\documentclass[12pt]{article}

\usepackage{amsmath,amssymb,amsfonts,amsbsy}
\usepackage{cite}
\usepackage{graphicx}
\usepackage{wrapfig}
\usepackage{epsfig}

\usepackage
{hyperref} 

\usepackage{times}
\usepackage{graphicx}
\usepackage{amssymb,amsmath}
\usepackage{fancybox}
\usepackage[latin1]{inputenc}
\usepackage[english]{babel}
\usepackage{umlaute}
\usepackage{pst-all}
\usepackage{epsfig}
\usepackage{rotating}

\usepackage{bbm} 
\usepackage{bm} 
\usepackage{color}                                                       %
\usepackage{dsfont} 
\usepackage{latexsym} 
\usepackage{lscape} 
\usepackage{mathrsfs} 
\usepackage{morefloats} 
\usepackage{slashed} 
\usepackage{psfrag}

\DeclareFontFamily{OT1}{mygreek}{}%
\DeclareFontShape{OT1}{mygreek}{m}{n}{<->omsegr}{}%
\DeclareFontShape{OT1}{mygreek}{b}{n}{<->omsegrb}{}%
\DeclareFontShape{OT1}{mygreek}{m}{it}{<->omsegri}{}%
\DeclareFontShape{OT1}{mygreek}{bx}{n}{<->sub * mygreek/b/n}{}%
\DeclareFontShape{OT1}{mygreek}{m}{sl}{<->sub * mygreek/m/it}{}%
\DeclareSymbolFont{Greekrm}{OT1}{mygreek}{m}{n}
\DeclareSymbolFont{Greekbf}{OT1}{mygreek}{b}{n}
\DeclareSymbolFont{Greekit}{OT1}{mygreek}{m}{it}
\DeclareMathSymbol{\omegab}{\mathalpha}{Greekbf}{119}

\def\a{\alpha}
\def\b{\beta}
\def\g{\gamma}
\def\vt{\vartheta}
\def\d{\delta}

\def\vt{\vartheta}

\def\sq2{\sqrt{\frac{\varepsilon_0}{\mu_0}} }

\newcommand{\be}{\begin{equation}}
\newcommand{\ee}{\end{equation}}
\newcommand{\bea}{\begin{eqnarray}}
\newcommand{\eea}{\end{eqnarray}}
\newcommand{\bdm}{\begin{displaymath}}
\newcommand{\edm}{\end{displaymath}}

\definecolor{wichtig}{rgb}{1,0,0} 
\definecolor{folge}{rgb}{0,0,1} 
\definecolor{liste}{rgb}{0,0.7,0} 

\definecolor{dark-green}{rgb}{0,0.7,0}
\definecolor{dark-blue}{rgb}{0,0.2,0.5}
\definecolor{med-blue}{rgb}{0,0.7,1}
\definecolor{mblue}{rgb}{0,0.2,1}
\definecolor{cnc}{rgb}{0.8,0,0}
\definecolor{light-red}{rgb}{1,0.8,0.8}
\definecolor{dark-yellow}{rgb}{1,0.8,0}
\definecolor{light-blue}{rgb}{0.8,0.9,1}
\definecolor{verylight-blue}{rgb}{0.93,0.95,1}
\definecolor{light-yellow}{rgb}{1,0.9,0.8}
\definecolor{grey}{gray}{0.88}

\def\ho{^}

\begin{document}

\setcounter{section}{0}
\setcounter{subsection}{0}
\setcounter{equation}{0}
\setcounter{figure}{0}
\setcounter{footnote}{0}
\setcounter{table}{0}

\begin{center}
  \textbf{ON ENERGY-MOMENTUM AND SPIN/HELICITY\\ OF QUARK AND GLUON
    FIELDS\footnote{Invited talk delivered at the XV Workshop on High
      Energy Spin Physics `DSPIN-13' in Dubna, Russia, 08--12 October
      2013 (file {\it DubnaSpin2013$\underline{\hspace{4pt}}$6.tex,} 02
        Feb.\ 2014).} 
}

\vspace{5mm}

Friedrich W.\ Hehl

\vspace{5mm}

\begin{small}
\emph{Institute for Theoretical Physics, University of
    Cologne, 50923 K\"oln, Germany} \\
  \emph{and Dept.\ of Physics \& Astronomy, University of
    Missouri, Columbia, MO 65211, USA} \\
\emph{E-mail: hehl@thp.uni-koeln.de}
\end{small}
\end{center}

\vspace{0.0mm} 

\begin{abstract}
  In special relativity, quantum matter can be classified according to
  mass-energy and spin. The corresponding field-theoretical notions
  are the {\it energy-momentum}-stress tensor ${\mathfrak T}$ and the
  {\it spin} angular momentum tensor ${\mathfrak S}$. Since each
  object in physics carries energy and, if fermionic, also spin, the
  notions of ${\mathfrak T}$ and ${\mathfrak S}$ can be spotted in all
  domains of physics. We discuss the $\mathfrak T$ and $\mathfrak S$
  currents in Special Relativity (SR), in General Relativity (GR) ,
  and in the Einstein-Cartan theory of gravity (EC). We collect our
  results in 4 theses: (i) The quark energy-momentum and the quark
  spin are described correctly by the canonical (Noether) currents
  $\mathfrak T$ and $\mathfrak S$, respectively. (ii) The gluon
  energy-momentum current is described correctly by the (symmetric and
  gauge invariant) Minkowski type current. Its (Lorentz) spin current
  vanishes, $\mathfrak{S}=0$. However, it carries helicity of plus or
  minus one.  (iii) GR contradicts thesis (i), but is compatible with
  thesis (ii). (iv) Within the viable EC-theory, our theses (i) and
  (ii) are fulfilled and, thus, we favor this gravitational theory.
\end{abstract}

\section{Introduction}

The {\it nucleon spin} and how it is built up in terms of spin and
orbital angular momentum contributions of the quark and gluon fields
is still under discussion. Recently, in this context, the problem has
been addressed of the appropriate energy-momentum and spin tensors of
quark and gluon fields, see the review paper of Leader and Lorc\'e
\cite{LeaderLorce:2013}. They emphasize the importance of the
splitting of the angular momentum of the gluon field into orbital and
spin parts. However, since the energy-momentum and angular momentum
distributions of a field are interrelated via the {\it orbital}
angular momentum, the angular momentum question can only be answered
if the energy-momentum distribution is treated at the same time. This
is an expression of the {\it semi-}direct product structure of the
Poincar\'e group $P(1,3):=T(4)\rtimes SO(1,3)$; here $T(4)$ denotes
the translation group and $SO(1,3)$ the Lorentz group.

These facts are, of course, recognized by Leader and Lorc\'e
\cite{LeaderLorce:2013} perfectly well, as can be seen by their
discussion of the so-called Belinfante and the canonical
energy-momentum tensors of both, the gluon and the quark fields. Even
though they mention general relativity (GR) in this context, their
main arguments are taken {}from special-relativistic quantum field
theory. On the other side it is known---we only remind of Weyl's
verdict \cite{Weyl} that only {\it ``the process of variation to be
  applied to the metrical structure of the world, leads to a true
  definition of the energy''} of matter---that an appropriate
gravitational theory is obligatory in order to get a clear insight
into the energy-momentum distribution of matter.

Why is this true? In Newton's gravitational theory the mass density of
matter is the source of gravity; in GR, by an appropriate
generalization, it is the (symmetric) Hilbert energy-momentum tensor
$\ho{\text{Hi}}\mathfrak{t}_{ij}$, which is computed by variation of
the matter Lagrangian ${\cal L}$ with respect to the metric tensor
$g_{ij}$, namely $\ho{\text{Hi}}\mathfrak{t}_{ij}:=2\d {\cal L}/\d
g\ho{ij}$.  Consequently, we have to assume that the energy-momentum
distribution is, in the classical limit, a measurable quantity and
that this localized energy-momentum distribution, with its $10$
components, is the source of the gravitational field. {\it As long as
  we subscribe to GR,} the Hilbert energy-momentum tensor is the only
viable energy-momentum tensor of matter, a fact that is put into doubt
by Leader and Lorc\'e.\footnote{``...we feel that the fundamental
  versions are the canonical and the Belinfante ones, since they
  involve at least local fields...'', see \cite{LeaderLorce:2013},
  page 92.}

Teryaev \cite{Teryaev:1999} already pointed out that the
energy-momentum tensor of matter will play a decisive role at the
interface between quantum chromodynamics (QCD) and gravity, see also
\cite{Teryaev:2003ch}. He discussed the gravitational moments of Dirac
particles, as done earlier by Kobzarev and Okun \cite{Kobzarev:1962}
and by Hehl et al.\ \cite{DiracCurrents}.

Let us recall the eminent importance of the Poincar\'e group. Wigner's
mass-spin classification of elementary (or fundamental) particles
\cite{Wigner} is at the basis of the standard model of particle
physics, the quark and the gluon are particular examples of it. The
mass-spin classification, by means of a scalar and a vector quantity,
underlines the particle aspect of matter. The corresponding notions
for elementary fields in classical field theory, are the
energy-momentum current\footnote{We use, for energy-momentum and spin,
  the notions `tensor' and `current' synonymously.} and the spin
current. Thus, the mass-spin classification of matter is mirrored on
the field-theoretical side by the canonical (Noether) energy-momentum
current $\mathfrak{T}_i{}\ho{k}$, with $4\times 4$ components, and the
canonical (Noether) spin current
$\mathfrak{S}_{ij}{}^{k}=-\mathfrak{S}_{ji}{}^{k}$ with its $6\times
4$ components. The relation of the Hilbert and the Noether
energy-momentum currents will be discussed further down.

On the gravitational side, there took some developments place that are
not without implications for the understanding of the energy-momentum
and the spin distribution of matter fields, see also the thermodynamic
considerations of Becattini \& Tinti \cite{Becattini:2011ev}. GR got a
competitor in the Einstein-Cartan(-Sciama-Kibble) theory of
gravitation (EC) or, more generally, in the Poincar\'e gauge theory of
gravitation (PG). A short outline and the classical papers of the
subject can be found in Blagojevi\'c and Hehl \cite{Reader}, see also
the review paper \cite{Erice95}.

The EC is a viable gravitational theory that can be distinguished {}from
GR at very high densities or at very small distances occurring in early
cosmology. The critical distance is $\ell_{\text{EC}}\approx
(\lambda_{\text{Co}}\ell_{\text{Pl}}\ho{2})\ho{1/3}$, with the Compton
wave length $\lambda_{\text{Co}}$ of the particle involved, about
$10\ho{-26}$ cm for the nucleon, and the Planck length
$\ell_{\text{Pl}}\approx 10\ho{-33}$ cm. Mukhanov \cite{Mukhanov} has
argued the the data of the Planck satellite support GR up to distances
of the order of $10\ho{-27}$ cm, that is, the same order of magnitude
where the deviations of EC are supposed to set in.

The EC-theory is a simple case of a PG-theory. The PG-theory is
formulated in a Riemann-Cartan (RC) spacetime with torsion
$C_{ij}{}^k$ ($=-C_{ji}{}^k$) and curvature $R_{ij}{}^{kl}$
($=-R_{ji}{}^{kl}=-R_{ij}{}^{lk}$). The gravitational Lagrangian of
PG-theory is, in general, quadratic in the field strengths torsion and
curvature. EC-theory is the simplest case, when the Lagrangian, apart
{}from the cosmological term, consists only of a linear curvature
piece $\sim R_{ij}{}^{ji}$ (summation!), the Riemann-Cartan
generalization of the Hilbert-Einstein Lagrangian. Then, additionally
to the gravitational effects of GR, we find a very weak
spin-spin-contact interaction that is governed by Einstein's
gravitational constant. But what is more relevant in the present
context is that in PG-theory---hence also in EC-theory---the source of
the Newton-Einstein type gravity is the {\it canonical
  energy-momentum} and the source of a Yang-Mills type strong gravity
the {\it canonical spin}.

However, one has to be careful in the details: Gauge field, like the
electromagnetic or the gluon field, do {\it not }carry canonical
(Lorentz) spin, but rather only {\it helicity}, see \cite{Leader}. In
this case, the canonical energy-momentum turns out to be what is
conventionally called the symmetrized energy-momentum. This will be
explained in detail.  With these provisos in mind, we can state that
the canonical tensors for energy-momentum and spin play the role of
sources of gravity in the PG-theory. Here we have an interface between
gravity and hadron physics as stressed by Teryaev \cite{Teryaev:1999}.

\section{Action principle, translational invariance}

We consider classical matter field $\Psi(x)$ (scalar, Weyl, Dirac,
Maxwell, Proca, Rarita-Schwinger, Fierz-Pauli etc.) in special
relativity (SR). The Minkowski spacetime $M_4$, with Cartesian
coordinates $x^i$ ($i,j,\dots=0,1,2,3$), carries a Lorentz metric
$g_{ij}\stackrel{*}{=} o_{ij}$ $:= {\rm diag}(+---)$. An isolated
material system with first order action $W_{\text{mat}}:=\frac 1c\int
d\Omega {\cal L}(\Psi,\partial\Psi)$ (see \cite{LL,Corson}) is
invariant under 4 translations, $x^{'i}=x^i+a^i$. The Noether theorem and
${\d{\cal L}}/{\d\Psi}=0$ yield the energy-momentum conservation in the form
\begin{eqnarray}\label{Tcan}
  \boxed{  \partial_j\mathfrak{T}_i{}^{\,j}=0}\,,
  \qquad\underbrace{\mathfrak{T}_i{}^{\,j}}_
  {4\times 4}:=\frac{\partial {\cal L}}{\partial
    \partial_j\Psi}\partial_i\Psi-{\cal L}\d_i^j\,,
\end{eqnarray}
with the canonical (Noether) energy-momentum tensor of type $
\left(\begin{smallmatrix}1\\1\end{smallmatrix}\right)$, also called
momentum current density. It is, in general, asymmetric and has 16
independent components.

With metric we can lower the upper index of $\mathfrak{T}_i{}^j$and
can decompose $\mathfrak{T}_{ij}$ irreducibly with respect to the
Lorentz group\footnote{The {Bach parentheses are $(ij):=\frac
    12\{ij+ji\} ,\,[ij]:= \frac 12\{ij-ji\}$, see Schouten
    \cite{SchoutenRicci}.}} (here ${\not\hspace{-2pt}
  \mathfrak{T}}_{ij}:= \mathfrak{T}_{(ij)}-\frac 14
g_{ij}\mathfrak{T}_{k}{}^k$):\vspace{-20pt}

\begin{eqnarray}
  \mathfrak{T}_{ij}&=& \hspace{38pt} \not\hspace{-2pt}
  \mathfrak{T}_{ij}\hspace{26pt}+\hspace{20pt}\mathfrak{T}_{[ij]}
  \hspace{20pt}+\hspace{13pt}\frac 14
  g_{ij}\mathfrak{T}_{k}{}^k\,,\hspace{4pt}\\
  16 &=& 9 \,(\text{sym.tracefree})\oplus\,6
  (\text{antisym.})\,\oplus\, \hspace{2pt} 1\, (\text{trace})\,.\nonumber
\end{eqnarray} 

An ansatz for a simple classical fluid (``dust'') is
\begin{equation}\label{dust}
\underbrace{\mathfrak{T}_i{}^{\,j}}_{\text{mom.\ curr.\
    d.}}=\underbrace{\mathfrak{p} _i}_{\text{mom.\ d.}}
\underbrace{u^{\,j}}_{\text{velocity}}\,\qquad\text{(observe natural index
  positions)}\,.
\end{equation}
If the momentum density is transported in the direction of the
velocity, $\mathfrak{p}_i={\rho} g_{ik}u^k$, with $\rho$ as
mass-energy density, then $\mathfrak{T}_{[ij]}=0$. A bit more refined
is the classical ideal (perfect, Euler) fluid, with $p$ as pressure:
\begin{equation}
\mathfrak{T}_{ij}=(\rho+p) u_iu_j -p g_{ij}\,,\qquad
  \mathfrak{T}_{[ij]}=0\,,\qquad \mathfrak{T}_{k}{}^k=\rho -3p\,.
\end{equation}

Superfluid $^3$He in the A-phase is a spin fluid of the convective
type, see Eq.(\ref{spinfluid}) below. The angular momentum law, as
formulated for the A-phase on p.\ 427 of Vollhardt \& W\"olfle
\cite{Vollhardt}, is a proof of this stipulation. This is an
irrefutable result that asymmetric stress tensors do exist in nature,
a fact doubted in many texts.

The quark current, as spin 1/2 current, should be of a similar type as
the superfluid $^3$He in the A-phase.  That is, the (physically
correct) energy-momentum current of the quark field should be
asymmetric and most probably the canonical (Noether) current
${\mathfrak T}_i{}^j$ of Eq.(\ref{Tcan}).

In {\it electromagnetism,} only $ {\not\hspace{-2pt}
  \mathfrak{T}}_{ij}$ survives (9 components), since it is massless,
that is, $\mathfrak{T}_{k}{}^k=0$, and carries {\it helicity}, but no
(Lorentz) spin, i.e., $\mathfrak{T}_{[ij]}= 0$. The analogous should
be true for the gluon field, since, like the Maxwell (photon) field,
it is a gauge field, see below for some more details.

Where took Einstein the symmetry of the energy-momentum tensor {}from?
Einstein, in \cite{EinsteinMeaning} on the pages 48 and 49, discussed
the symmetry of the energy-momentum tensor of Maxwell's
theory. Subsequently, on page 50, he argued: ``{\it We can hardly
  avoid making the assumption that in all other cases, also, the space
  distribution of energy is given by a symmetrical tensor,
  $T_{\mu\nu}$, ...}'' This is hardly a convincing argument if one
recalls that the Maxwell field is massless. As we saw, the A-phase of
$^{3}$He contradicts Einstein's assumption. Asymmetric energy-momentum
tensors are legitimate quantities in physics and, the symmetry of an
energy-momentum tensor has to retire as a generally valid rule.

\section{Lorentz invariance}

Invariance under 3+3 infinitesimal Lorentz transformations,
$x^{'i}=x^i+\omega^{ij}x_j$, with $\omega^{(ij)}=0$, yields, via the
Noether theorem and ${\d{\cal L}}/{\d\Psi}=0$, angular-momentum
conservation,
\begin{eqnarray}\label{angcon}
  \partial_k\big(\underbrace{\mathfrak{S}_{ij}{}^k}_{\text{spin}}+\underbrace{
    x_{[i} \mathfrak{T}_{j]}{}^k}_{\text{orb.\ angular mom.}}\big)=0\,,
  \qquad\underbrace{\mathfrak{S}_{ij}{}^k}_{6\times
    4}:=-\frac{\partial{\cal L}}{\partial\partial_k\Psi}
  \hspace{-5pt}\underbrace{f_{ij}}_{\text{ Lor.\ gen.}}\hspace{-5pt}\Psi
  =-\mathfrak{S}_{ji}{}^k\,.
\end{eqnarray}
The canonical (Noether) spin $\mathfrak{S}_{ij}{}^k$, the spin current
density, is a tensor of type $ \left(\begin{smallmatrix}1
    \\2\end{smallmatrix}\right)$, plays a role in the interpretation
of the Einstein-de Haas effect (1915) . If we differentiate in
(\ref{angcon})$_1$ the second term and apply
$\partial_k\mathfrak{T}_i{}^k=0$, then we find a form of angular
momentum conservation that can be generalized to curved and contorted
spacetimes ($x^i$ is not a vector in general):
\begin{equation}
  \partial_k\left(\mathfrak{S}^{ijk} +x^{[i}\mathfrak{T}^{j]k}\right)=0
  \qquad\Longrightarrow\qquad \boxed{ \partial_k\mathfrak{S}^{ijk}
    -\mathfrak{T}^{[ij]}=0}\,.
\end{equation}
If $\mathfrak{S}^{ijk}=0$, then $\mathfrak{T}^{[ij]}=0$, that is, the
energy-momentum tensor is symmetric, but not necessarily vice versa.

The irreducible decomposition, with the axial vector piece $
^{\text{AX}}\mathfrak{S}_{ijk} :=\mathfrak{S}_{[ijk]}$ and the vector
piece $^{\text{VEC}}\mathfrak{S}_{ij}{}^{k}:=\frac
23\mathfrak{S}_{[i|\ell}{}^\ell \d_{|j]}^k$, reads:
\begin{eqnarray}
\mathfrak{S}_{ij}{}^{k}&=&\,^{\text{TEN}}\mathfrak{S}_{ij}{}^{k}+
\,^{\text{VEC}}\mathfrak{S}_{ij}{}^{k}+\,^{\text{AX}}\mathfrak{S}_{ij}{}^{k}\,,\\
\nonumber
24&=&\hspace{10pt} 16\hspace{17pt}\oplus\hspace{18pt} 4\hspace{18pt}
\oplus\hspace{15pt} 4\,,
\end{eqnarray}

The Weyssenhoff ansatz for a classical spin fluid is again of the
convective type. We take
\begin{equation}\label{spinfluid}
  \underbrace{\mathfrak{T}_i{}^{\,j}}_{\text{mom.\ curr.\
      d.}}=\underbrace{\mathfrak{p} _i}_{\text{mom.\ d.}}
  \underbrace{u^{\,j}}_{\text{velocity}}\,
  \qquad\text{and}\qquad\underbrace{\mathfrak{S}_{ij}{}^{k}}_{\text{spin curr.\
      d.}}=\underbrace{\mathfrak{s} _{ij}}_{\text{spin d.}}
  \underbrace{u^{k}}_{\text{velocity}}=-\mathfrak{S}_{ji}{}^{k}\,.
\end{equation}
The momentum density $\mathfrak{p} _i$ is no longer proportional to
the velocity, as it was in (\ref{dust}). Usually, the constraint
$\mathfrak{s}_{ij}u^j=0$ is assumed.

For the Dirac field, the spin current is totally antisymmetric,
$\stackrel{D}{\mathfrak{S}}_{ijk}= \stackrel{D}
{\mathfrak{S}}_{[ijk]}$. Thus, only the axial vector spin current
survives, ${} ^{\text{AX}}\!\stackrel{D}{\mathfrak{S}}_{ijk} \ne
0$. The Dirac field is highly symmetric. Accordingly, we can introduce
the spin flux {\it vector}
\begin{equation} 
  {\cal S}^i:={{\frac{1}{3!}}}\epsilon^{ijkl}\mathfrak{S}_{jkl}\quad\sim\quad
  (\text{spin flux density 1 comp., spin density 3 comps.})\,.
\end{equation}
The spin density distribution is spatially isotropic.

\section{Poincar\'e invariance}

We collect our results: The Poincar\'e invariance of the action yields
the $4+6$ conservation laws,
\begin{eqnarray}\label{P1}
  \partial_k\mathfrak{T}_i{}^k\hspace{25pt}&=& 0\qquad 
  \text{(energy-momentum conservation)}\,,\\ \label{P2}
 \partial_k\mathfrak{S}_{ij}{}^k
    -\mathfrak{T}_{[ij]}&=& 0\qquad \text{(angular momentum conservation)}\,.
\end{eqnarray}
These field theoretical notions $\mathfrak{T}_i{}^k$ and
$\mathfrak{S}_{ij}{}^k$ have their analogs in a the particle
description of matter. The Lie algebra of the Poincar\'e group reads
(see \cite{Tung} for details, $\hbar=1$):
\begin{align} [P_i,P_j] & =  0\,, \nonumber\\
  [J_{ij}, P_k] & = 2i\,g_{k[i}P_{j]}\qquad (\text{transl.\ and Lorentz
    transf.\ mix, as in }
  \mathfrak{S}_{ijk}+x_{[i}\mathfrak{T}_{j]k})\,,\\
  [J_{ij},J_{kl}]&= 2i\!\left(g_{k[i}J_{j]l}-g_{l[i}J_{j]k}\right)\,.\nonumber
\end{align}
We recognize its semidirect product structure, as it is manifest in
the existence of orbital angular momentum. The ``square roots'' of the
Casimir operators $P^2$ (mass square) and $W^2$ (spin square), with
the Pauli-Luba\'nski vector $W^i:=\frac 12\epsilon^{ijkl}\,J_{jk}P_l$,
correspond to $\mathfrak{T}_i{}^k$ and $\mathfrak{S}_{ij}{}^k$.

\section{Exterior calculus in a Riemann-Cartan (RC) space, the
  electromagnetic/gluon energy-momentum, and the Dirac field}

We introduce the generally covariant calculus of {\it exterior
  differential forms} that is valid not only in Minkowski space, but
also in the RC-spacetime of the Poincar\'e gauge theory of gravity,
see \cite{PRs}. We work with an {orthonormal coframe} (tetrad)
${\vt^\a}=e_i{}{}^\a dx^i$ and a {Lorentz connection}
${\Gamma^{\a\b}}=\Gamma_i{}^{\a\b}dx^i=-\Gamma^{\b\a}$;
the fields are exterior forms (0-forms, 1-forms,..., 4-forms) with
values in the algebra of some Lie group; the frame (or anholonomic)
indices are in Greek, $\a,\b,\dots=0,1,2,3$. The electromagnetic
potential is a 1-form $A=A_i dx^i$, the field strength a 2-for $F:=d
A=\frac 12 F_{ij}dx^i\wedge dx^j$, the exterior derivative is denoted
by $d$, the gauge covariant exterior derivative is by $D$, for details see
\cite{Birkbook}.

The matter currents translate {}from tensor to exterior calculus as
follows: Energy-momentum 3-form ${\mathfrak{T}_\a}
=\mathfrak{T}_\a{}^\g\, {}^\star\vt_\g=\d L_{\text{mat}}/\d {\vt^\a}$,
spin 3-form ${\mathfrak{S}_{\a\b}} =\mathfrak{S}_{\a\b}{}^\g\,
{}^\star\vt_\g=\d L_{\text{mat}}/\d {\Gamma^{\a\b}}$, with the Hodge
star $^\star$. Here we displayed already the variational expression,
which will be explained below.

Maxwell's vacuum field ${A(x)}$ is a 1-form, a geometrical object
independent of coordinates and frames. As such, it has {vanishing
  Lorentz-spin, $ {\mathfrak{S}_{\a\b}}=0$,} but {helicity $\pm
  1$}. The analogous is true for the {gluon} field. As a consequence,
in exterior calculus, its canonical (i.e.\ Noether) energy-momentum
3-form is symmetric and gauge invariant directly, see \cite{PRs},
footnote 53. Conventionally, see \cite{LL}, the {\it coordinate
  dependent} components $A_i$ of $A$ are used in the Lagrangian
formalism, see also the clarifying considerations of Benn et al.\
\cite{Benn}.\bigskip

\noindent{\bf Thesis 1:} {\it The energy-momentum current 3-form of
  the free gluon field} $F=DA$ {\it is given by the Minkowski type
  expression \cite{Minkowski:1908}}
\begin{equation}
  \mathfrak{T}_\a=\frac 12 [F\wedge (e_a\rfloor {}^\star\!
  F)-{}^\star\!F\wedge(e_\a\rfloor F)]\qquad\text{or}\qquad\mathfrak{T}_i{}^j
  =\frac 14\d_i^jF_{kl}F^{kl}-F_{ik}F^{jk}\,.
\end{equation}
{\it The (Lorentz) spin current of the gluon field vanishes,
  $\mathfrak{S}_{\a\b}{}^\g=0$, the gluon orbital angular momentum current
  is given by $x_{[\a}\mathfrak{T}_{\b]}$ and represents the total
  angular momentum. As a gauge potential, the gluon is described by 
a 1-form and has helicity $\pm 1$.} \bigskip

The second example, Dirac field in exterior calculus for illustration. Its Lagrangian reads,
\begin{equation}
L_{\rm D}=\frac i2(\overline{\Psi}{}^\star\!\gamma\wedge D\Psi+
\overline{D\Psi}\wedge\,^\star\!\gamma\Psi)+\,^\star
m\overline{\Psi}\Psi\,,
\end{equation} with $\gamma:=\gamma_\a\vt^\a$ and
$\gamma_{(\a}\gamma_{\b)}=o_{\a\b} \boldsymbol{1}_4$.  The 3-forms of
the canonical momentum and spin current densities are ($D_\a:=e_\a\rfloor
D$, here $\rfloor$ denotes the interior product sign):
\begin{eqnarray}{\mathfrak{T}_\a}{=}{\frac
    i2 (\overline{\Psi} \,^\star\!\gamma\wedge D_\a\Psi+
    \overline{D_\a\Psi}\wedge\,^\star\!\gamma\Psi)}\,,\label{DiracT} \qquad
  {\mathfrak{S}_{\a\b}} {=}
  {\frac 14\vt_\a\wedge\vt_\b\wedge \overline{\Psi}
    \gamma\gamma_5\Psi}\,.
\end{eqnarray}
In Ricci calculus $\mathfrak{S}_{\a\b\g}=\mathfrak{S}_{[\a\b\g]}=\frac
14 \epsilon_{\a\b\g\d} \overline{\Psi}\gamma_5\gamma^\d \Psi$. Because
of the equivalence principle, the {\it inertial currents}
${\mathfrak{T}_\a}$ and $ {\mathfrak{S}_{\a\b}}$ are, at the same
time, the gravitational currents of the classical Dirac field. A
decom\-position of $(\mathfrak{T}_\a,\mathfrak{S}_{\a\b})$ \`a la
Gordon, yields the {\it gravitational} moment densities of the Dirac
field \cite{DiracCurrents}; it is a special case of relocalization,
see below.\bigskip

\noindent{\bf Thesis 2:} {\it The canonical (Noether) energy-momentum and the
  canonical (Noether) spin current 3-forms of a Dirac/quark field are
  given by the expressions in Eq.(\ref{DiracT}).}

\section{Relocalization of energy-momentum and spin distribution}
We redefine the canonical currents ${\mathfrak{T}}{}_{i}{}^{j}$
and ${\mathfrak{S}}{}_{ij}{}^{k} $ by adding curls, see
\cite{ROMP,DiracCurrents},
\begin{eqnarray}\label{Phead}
  \widehat{\mathfrak{T}}{}_{i}{}^{j} := \,\mathfrak{T}_{i}{}^{j} 
  + \partial_k\,Y_{i}{}^{jk},\qquad
  \widehat{\mathfrak{S}}{}_{ij}{}^{k}:=
  \,\mathfrak{S}_{ij}{}^{k}
  +Y_{[ij]}{}^k+\partial_l\,Z_{ij}{}^{kl}\,,
\end{eqnarray}
with the arbitrary antisymmetric super-potentials $Y_{i}{}^{jk} =
-\,Y_{i}{}^{kj}$ and $ Z_{ij}{}^{kl} =-\,Z_{ij}{}^{lk}=
-\,Z_{ji}{}^{kl}$. We substitute (\ref{Phead})$_1$ and the partial
derivative of (\ref{Phead})$_2$ into (\ref{P1}) and (\ref{P2}), Then
we recognize that these {\it relocalized} currents fulfill the
original conservation laws:
\begin{eqnarray}
 \partial_j\widehat{\mathfrak{T}}_i{}^j=0\,,\qquad
 \partial_k\widehat{\mathfrak{S}}_{ij}{}^k
    -\widehat{\mathfrak{T}}_{[ij]}= 0\,.
\end{eqnarray}
The integrated {\it total} energy-momentum and the {\it total} angular
momentum of an insular material system are invariant under
relocalization \cite{ROMP}. However, ``relocalization invariance''
under the transformations specified in (\ref{Phead}) is {\it not} a
generally valid physical principle. It should rather be understood as
a formal trick to compute the total energy-momentum and angular
momentum in a most convenient way.

It is convenient to introduce a new superpotential $U$ that is
equivalent to $Y$ by 
\begin{equation}\label{NewU}
  U_{ij}{}^k:=-Y_{[ij]}{}^k=-U_{ji}{}^k\qquad\Longrightarrow \qquad  
  Y_i{}^{jk}=-U_{i}{}^{jk}+U^{jk}{}_i-U^{k\,.\,j}_{\hspace{4pt}i}\,.
\end{equation}

The {\it Belinfante} relocalization (1939) is a special case:
Belinfante \cite{Belinfante} effectively required
$\widehat{\mathfrak{S}}{}_{kl}{}^{j} =0$. Then, by (\ref{Phead})$_2$
and (\ref{NewU})$_1$, $\mathfrak{S}_{ij}{}^{k}
=U_{ij}{}^k-\partial_l\,Z_{ij}{}^{kl}$ and the relocalized
energy-momentum, $^{\text{Bel}}\mathfrak{t}_{i}{}^{j} :=
\widehat{\mathfrak{T}}{}_{i}{}^{j}$, with
$\widehat{\mathfrak{S}}{}_{kl}{}^{j} =0$, reads
\begin{eqnarray} ^{\text{Bel}}\mathfrak{t}_{i}{}^{j} =
  \mathfrak{T}_i{}^j-\partial_k\!\left( \mathfrak{S}_{i}{}^{jk}-\mathfrak{S}^{jk}{}_i+\mathfrak{S}{}^{k\,.\,j}_{\hspace{4pt}i}
   \right)\quad\text{with}\quad
  \boxed{\partial_{j}{} ^{\text{Bel}}\mathfrak{t}_{i}{}^{j}={}0,\qquad
    ^{\text{Bel}} \mathfrak{t}_{[kl]} = 0\,.}
\end{eqnarray}

For the Dirac field, because of the total antisymmetry of
$\mathfrak{S}_{ijk}$, we find simply $^{\text{Bel}}\mathfrak{t}_{ij}=
\,^{\text{Bel}}\mathfrak{t}_{(ij)}=\mathfrak{T}_{(ij)}$, see
\cite{Tetrode}. Incidentally, the Gordon relocalization, mentioned
above, {\it differs} {}from the Belinfante relocalization.

\section{Dynamic Hilbert energy-momentum in general relativity}

How can we choose amongst the multitude of the relocalized
energy-momentum tensors and spin tensors? After all, as physicists we
are convinced that the energy and the spin distribution of matter (but
not of gravity!) are observable quantities, at least in the classical
domain. There must exist physically correct and unique energy-momentum
and spin tensors in nature. The Belinfante recipe was to kill
$\mathfrak{T}_{[kl]}$ in order to tailor the energy-momentum for the
application in Einstein's field equation. 

Already in 1915, Hilbert defined the dynamic energy-momentum as the
response of the matter Lagrangian to the variation of the metric
\cite{Hilbert}:
\begin{equation}\label{HilbertEM}
^{\rm Hi}\mathfrak{t}_{ij}:=2{\d \mathfrak{L}_{\text{mat}}(g,\Psi\,,
  \stackrel{\{\}}{\nabla}\Psi)}/{\d g^{ij}}\,;
\end{equation} $g^{ij}$ (or its
reciprocal $g_{kl}$) is the gravitational potential in GR. The matter Lagrangian
is supposed to be {\it minimally coupled} to $g^{ij}$, in accordance
with the equivalence principle. Only in gravitational theory, in
which spacetime can be deformed, we find a real local definition of
the material energy-momentum tensor. The Hilbert definition is
analogous to the relation {}from elasticity theory
``$\text{stress} \sim\d \text{(elastic energy)}/\d
\text{({strain})}$''. Recall that strain is defined as 
$\varepsilon^{ab}:=\frac 12\left(^{\rm (defo)}g^{ab}-\,^{\rm
  (undefo)}g^{ab}\right)$, see \cite{LLelast}. Even the factor 2 is reflected 
in the Hilbert formula.

Rosenfeld (1940) has shown \cite{Rosenfeld:1940}, via Noether type
theorems, that the Belinfante tensor $^{\text{Bel}}\mathfrak{t}_{ij}$,
derived within SR, coincides with the Hilbert tensor $^{\rm
  Hi}\mathfrak{t}_{ij}$ of GR. Thus, the Belinfante-Rosenfeld recipe
yields...  \medskip

\noindent {\bf Thesis 3:} {\it In the framework of GR, the Hilbert
  energy-momentum tensor}
\begin{equation}\label{HiBel}
^{\rm Hi}\mathfrak{t}_{i}{}^j={}^{\text{Bel}}\mathfrak{t}_{i}{}^j= 
\mathfrak{T}_i{}^j-\nabla_k\!\left(\mathfrak{S}_{i}{}^{jk}-\mathfrak{S}^{jk}{}_i
  +\mathfrak{S}^{k\,.\,i}_{\hspace{4pt}i}\right)=\,^{\rm Hi}\mathfrak{t}^j{}_i\,,
\end{equation}
{\it localizes the energy-momentum distribution correctly; here $(
  \mathfrak{T}_i{}^j,\mathfrak{S}_{ij}{}^{k})$ are the canonical Noether
  currents. The spin tensor attached to $^{\rm Hi}\mathfrak{t}_{i}{}^j$ vanishes.}\medskip

The Rosenfeld formula (\ref{HiBel}) identifies the Belinfante with the
Hilbert tensor. In other words, the Belinfante tensor provides the
correct source for Einstein's field equation.  As long as we accept GR
as the correct theory of gravity, the localization of energy-momentum
and spin of matter is solved. This state of mind is conventionally
kept till today by most theoretical physicists. In passing, one should
note that the spin of matter has a rather auxiliary function in this
approach. After all, the spin of the Hilbert-Belinfante-Rosenfeld
tensor simply vanishes.

However, the Poincar\'e gauge theory of gravity (PG; Sciama, Kibble
1961, see \cite{Reader} for a review), in particular the viable
Einstein-Cartan theory (EC) with the curvature scalar as
gravitational Lagrangian, has turned the Rosenfeld formula
(\ref{HiBel}) upside down.

\section{Dynamic Sciama-Kibble spin in Poincar\'e gauge theory}

The gauging of the Poincar\'e group identifies as gauge potentials the
orthonormal coframe $\vt^\a=e_i{}^\a dx^i$ and the Lorentz connection
$\Gamma^{\a\b}=\Gamma_i{}^{\a\b}dx^i=-\Gamma^{\b\a}$. The spacetime
arena of the emerging Poincar\'e gauge theory of gravity (PG) is a
Riemann-Cartan space with Cartan's {\it torsion} and with
Riemann-Cartan {\it curvature} as gauge field strength, respectively
\cite{Erice95}:
\begin{equation}
C_{ij}{}^\a:=\nabla_{[i} e_{j]}{}^\a,\; R_{ij}{}^{\a\b}:=
\text{``}\nabla\text{''}\!_{[i}\Gamma_{j]}{}^{\a\b}\quad\quad
(or\;C^\a=D\vt^\a,\; R^{\a\b}=\text{``}D\text{''}\Gamma^{\a\b}).
\end{equation}
The energy-momentum and angular momentum laws generalize to 
\begin{eqnarray}
  \stackrel{*}{\nabla}_k\mathfrak{T}_i{}^k=
  \underbrace{C_{ik}{}^{\ell}}_{\text{torsion}}\mathfrak{T}_\ell{}^k+
\underbrace{R_{ik}{}^{lm}}_{\text{curvature}} \mathfrak{S}_{lm}{}^k\,,\qquad
  \stackrel{*}{\nabla}_k\mathfrak{S}_{ij}{}^k
  -\mathfrak{T}_{[ij]}= 0\,;
\end{eqnarray}
here $\stackrel{*}{\nabla}_k:={\nabla}_k+C_{k\ell}{}^\ell$. GR is the
subcase for $\mathfrak{S}_{ij}{}^k=0$, see also the reviews
\cite{LompayPetrov,Lompay}.  The material currents are defined by
variations with respect to the potentials:
\begin{equation}^{\rm SK}\mathfrak{T}_{\a}{}^i={\d
    \mathfrak{L}_{\text{mat}}(e,\Gamma,\Psi\,,
  \stackrel{\Gamma}{D}\Psi)}/{\d e_i{}^\a},\qquad{
  ^{\rm SK}\mathfrak{S}_{\a\b}{}^i}={\d
  \mathfrak{L}_{\text{mat}}(e,\Gamma,\Psi\,,
  \stackrel{\Gamma}{D}\Psi)}/{\d \Gamma_i{}^{\a\b}}\,.
\end{equation}
This Sciama-Kibble definition of the spin (1961) is only possible in
the Riemann-Cartan spacetime of PG. It is analogous to the relation
``${\text{moment\, stress}} \sim \d \text{(elastic energy)}/\d
\text{({contortion})}$'' in a Cosserat type medium, the contortion being
a ``rotational strain'', see \cite{HehlObukhovEssay}.

The application of the Lagrange-Noether machinery to the minimally
coupled action function yields, after a lot of algebra, the final
result, see \cite{PRs}:
\begin{equation} ^{\rm SK}\mathfrak{T}_{\a}{}^i=\mathfrak{T}_{\a}{}^i\,,\qquad
^{\rm SK}\mathfrak{S}_{\a\b}{}^i=\mathfrak{S}_{\a\b}{}^i\,.
\end{equation} 
The dynamically defined energy-momentum and spin currents \`a la
Sciama-Kibble coincide with the canonical Noether currents of
classical field theory. \bigskip

\noindent{\bf Thesis 4:} {\it Within PG, the quark energy-momentum and
  the quark spin are distributed in accordance with the canonical
  Noether currents $\mathfrak{T}_\a{}^i$ and $\mathfrak{S}_{\a\b}{}^i$,
  respectively.}\medskip

\noindent This is in marked contrast to the doctrine in the context of
GR.  

We express the canonical energy-momentum tensor in terms of the Hilbert
one (see \cite{Kopczynski:1989}):\vspace{-5pt}
\begin{eqnarray}\label{RosenfeldNew}
  ^{\rm SK}\mathfrak{T}_{\a}{}^i=\mathfrak{T}_{\a}{}^i= {} ^{\rm
    Hi}\mathfrak{t}_{\a}{}^i+\stackrel{*}{\nabla}_k\!(\mathfrak{S}_{\a}{}^{ik}
  -\mathfrak{S}^{ik}{}_{\a}+\mathfrak{S}^{k\hspace{4pt}i}_{\;\a}) \,,\qquad
  ^{\rm SK}\mathfrak{S}_{\a\b}{}^i=\mathfrak{S}_{\a\b}{}^i\,.
\end{eqnarray}
The new Rosenfeld formula (\ref{RosenfeldNew})$_1$ reverses its
original meaning in (\ref{HiBel}). Within PG, the canonical tensor
$\mathfrak{T}_{\a}{}^i$ represents the correct energy-momentum
distribution of matter and the (sym)met\-ric Hilbert tensor now plays
an auxiliary role. In GR, it is the other way round. Moreover, we are
now provided with a dynamic definition of the canonical spin
tensor. In GR, the spin was only a {\it kinematic} quantity floating
around freely.

These results on the correct distribution of material energy-momentum
and spin in the framework of PG are are {\it independent} of a
specific choice of the {\it gravitational} Lagrangian.  However, if we
choose the RC curvature scalar as a gravitational Lagrangian, we
arrive at the Einstein-Cartan(-Sciama-Kibble) theory of gravitation,
which is a {viable} theory of gravity competing with
GR.

\section{An algebra of the momentum and the spin currents?} 

We discussed exclusively classical field theory. Can we learn
something for a corresponding quantization of gravity? Our classical
analysis has led us to the gravitational currents $\mathfrak{T}_\a$
and $\mathfrak{S}_{\a\b}$. They represent the sources of gravity.

In strong and in electroweak interaction, before the standard model
had been worked out, one started with the current algebra of the
phenomenologically known {\it strong} and the {\it electroweak}
currents (see Sakurai \cite{Sakurai}, Fritzsch et al.\ \cite{Fritzsch},
and also Cao \cite{Cao}).

Schwinger (1963) studied, for example, the equal time commutators of
the components of the Hilbert energy-momentum tensor
\cite{Schwinger}. Should one try to include also the spin tensor
components and turn to the canonical tensors?

In the Sugawara model (1968), ``{\it A field theory of currents}'' was
proposed \cite{Sugawara:1968} with 8 vector and 8 axial vector
currents for strong interaction and a symmetric energy-momentum
current for gravity that was expressed bilinearly in terms of the
axial and the vector currents. Now, when we have good arguments that
the gravitational currents are $\mathfrak{T}_\a$ and
$\mathfrak{S}_{\a\b}$, one may want to develop a corresponding current
algebra by determining the equal time commutator of these
currents....

\section{Acknowledgments} 
I'd like to thank Anatoly Efremov, Chair of DSPIN-13, and Oleg Teryaev
for the invitation and for their hospitality in Dubna. For
discussions I am most grateful to Francesco Becattini (Frankfurt
\& Florence), Yuri Obukhov (Moscow), Oleg Teryaev (Dubna), and
Sasha Silenko (Dubna \& Minsk). I acknowledge support by the
German-Russian Heisenberg-Landau program.


\end{document}